\documentclass[superscriptaddress, twocolumn]{revtex4-2}

\usepackage{amsmath}
\usepackage{graphicx}
\usepackage{hyperref}

\begin{document}

\title{Reduction of hydroxyl groups in optical nanofibers via in-fiber laser heating}
\author{Koki Oguri}
\affiliation{Department of Applied Physics, Waseda University, 3-4-1 Okubo, Shinjuku, Tokyo 169-8555, Japan}
\author{Ken-ichi Harada}
\affiliation{Department of Applied Physics, Waseda University, 3-4-1 Okubo, Shinjuku, Tokyo 169-8555, Japan}
\author{Samuel K. Ruddell}
\affiliation{Department of Applied Physics, Waseda University, 3-4-1 Okubo, Shinjuku, Tokyo 169-8555, Japan}
\author{Karen E. Webb}
\affiliation{Department of Applied Physics, Waseda University, 3-4-1 Okubo, Shinjuku, Tokyo 169-8555, Japan}
\author{Takao Aoki}
\email{E-mail: takao@waseda.jp}
\affiliation{Department of Applied Physics, Waseda University, 3-4-1 Okubo, Shinjuku, Tokyo 169-8555, Japan}
\affiliation{RIKEN Center for Quantum Computing (RQC), Wako, Saitama 351-0198, Japan}

\begin{abstract}
Optical nanofibers fabricated using a standard oxyhydrogen flame exhibit optical losses at wavelengths around 1385~nm due to absorption by embedded hydroxyl groups, posing a challenge for the realization of quantum electrodynamics systems using ytterbium atoms. Here, we establish a method for the reduction of hydroxyl groups by heating them with a laser guided within the optical nanofiber under vacuum conditions. The temperature of the optical nanofiber during heating is estimated by monitoring the phase shift of the transmitted light using an interferometer. As a result, evidence suggesting that hydroxyl groups were desorbed by laser heating was obtained.
\end{abstract}

\maketitle

\noindent Cavity quantum electrodynamics (CQED) systems enhance atom--light interactions by strongly confining light within optical cavities, and present a promising platform for the realization of quantum communication and distributed quantum computation networks~\cite{Kimble2008, Webb2026}.
Combining atoms with high-quality optical cavities can allow for the manipulation of pure quantum states, enabling deterministic processing and storage of quantum information. Additionally, the use of photons as carriers of quantum information naturally lends itself to fully networked platforms, where entanglement is distributed over spatially delocalized quantum processing nodes~\cite{Reiserer2015, Ritter2012, Covey2023}.
Optical nanofiber~(ONF) cavities, formed by fiber Bragg grating mirrors enclosing a section of fiber tapered to a diameter comparable to the wavelength of guided light, show particular promise~\cite{lekien2009, Ruddell2020, Tanaka2026, Horikawa2024, Webb2026}. Here, the evanescent field of the nanofiber can provide a clean interface between light and a quantum emitter located near to the nanofiber surface, with the all-fiber nature of the cavity allowing for seamless integration into optical fiber communication channels~\cite{Kato2015, Kato2019, White2019}. 
Additionally, the fabrication of ONF cavities with finesse values greater than 3000 have been demonstrated, corresponding to projected single-atom cooperativities exceeding those of conventional free-space Fabry-P\'erot cavities~\cite{Horikawa2024}.
Among the various atomic species, ytterbium is considered a promising candidate for CQED systems~\cite{sunami2025} due to its long coherence time~\cite{jenkins2022} and rich energy-level structure in the telecom-band. In particular, strong atom--photon coupling is predicted when using the 1389~nm transition of $^{171}$Yb with state-of-the art nanofiber Fabry-P\'erot cavities~\cite{sunami2025}. The use of telecom transitions also enables direct compatibility with existing optical-fiber communication networks for long-distance quantum communication~\cite{li2024, li2025}.

However, when optical nanofibers are fabricated using the flame-brush method~\cite{birks1992} with a standard oxygen-hydrogen flame, water molecules produced during the process react with siloxane bonds (Si–O–Si) near the silica glass surface, forming silanol groups (Si–OH)~\cite{walrafen1978, stone1982, yokomachi1987, bredol1990, humbach1996, rose1997, rose1998, plotnichenko2000, yu2016}. These hydroxyl groups give rise to a wide absorption band centered at approximately 1385~nm, corresponding to the first overtone of the O–H stretching vibration~\cite{stone1982, yokomachi1987, bredol1990, humbach1996, rose1997, rose1998, plotnichenko2000, yu2016}. 
As a large fraction of the optical field propagates outside of the ONF in the form of an evanescent field, absorption by OH groups within the nanofiber results in significant optical loss.

A previous study has proposed fabricating ONFs using a deuterium–oxygen flame to suppress water-related absorption at 1389~nm~\cite{Horikawa2025}. However, deuterium gas is considerably more expensive than hydrogen gas, increasing the practical barrier to its implementation. Therefore, in this work, we propose an alternative approach in which nanofibers are fabricated using a conventional oxyhydrogen flame and are subsequently treated by laser heating to reduce OH~groups introduced during fabrication.
The temperature of the optical nanofiber during laser heating is estimated from the measured phase shift of a laser guided within the nanofiber, while incorporation and desorption of OH~groups are monitored from changes in absorption loss in the transmission spectrum near 1385~nm. 
While temperature-dependent changes in OH absorption bands have previously been reported in optical fibers\cite{yu2016}, to the best of our knowledge, no study has reported a reduction in the absorption band near 1385~nm induced by laser heating of an ONF under vacuum conditions. 

\begin{figure}[hbt!]
\centering
\includegraphics[width=\columnwidth]{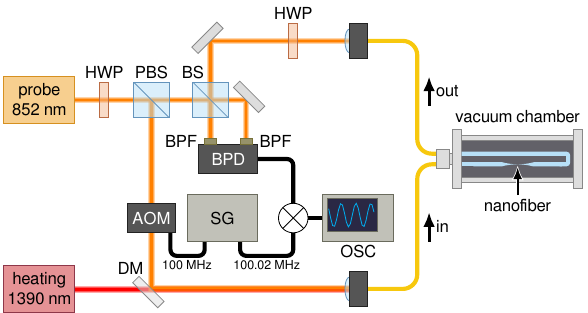}
\caption{Experimental setup used for measuring the phase shift of the ONF during laser heating. HWP: half-wave plate, PBS: polarizing beam splitter, BS: beam splitter, BPF: band-pass filter, AOM: acousto-optic modulator, DM: dichroic mirror, SG: signal generator, OSC: oscilloscope.}
\label{fig:setup}
\end{figure}
The ONFs used in this study were fabricated from a low-OH optical fiber (SMF-28 Ultra, Thorlabs), and tapered using the flame-brush method with an oxyhydrogen flame~\cite{birks1992, Ruddell2020}, adding hydroxyl groups to the nanofiber during the process. The waist diameter and length of the final ONF were approximately 500~nm and 5~mm, respectively. Following fabrication, the nanofiber was glued onto an Invar jig using vacuum-compatible UV glue (Dymax OP-29-GEL) and installed inside a metal vacuum chamber, which was then pumped down to below $10^{-4}$~Pa to allow for sufficient heating by the heating laser and prevent additional water absorption.

In order to monitor the ONF temperature, we utilize the optical heterodyne setup shown in Fig.~\ref{fig:setup}. Light from an 852~nm distributed Bragg reflector~(DBR) laser (DBR852PN, Thorlabs) is split by a polarizing beam splitter~(PBS), with a 100~MHz frequency shift applied to one output by an acousto-optic modulator~(AOM), before passing through the ONF. This light then interferes with the unshifted light from the second PBS output, and the resulting beat signal is detected with a balanced photodetector~(BPD). 
The signal from the BPD is mixed with a 100.02~MHz reference frequency from a signal generator~(SG), where the resulting 20~kHz intermediate-frequency signal allows us to extract the relative phase shift between the two beams, corresponding to the change in optical path length of the ONF. The temperature of the nanofiber can then be extracted using a thermal model~\cite{Anderson2018, wuttke2013}, which is discussed below.
For laser heating, a distributed-feedback~(DFB) single-frequency laser (1390LD-2-0-0, AeroDIODE) with a center wavelength of 1390~nm was inserted into the ONF along with the 852~nm DBR laser using a dichroic mirror~(DM). By employing a laser near 1385 nm, we are able to target the first overtone of O–H stretching~\cite{stone1982, yokomachi1987, bredol1990, humbach1996, rose1997, rose1998, plotnichenko2000, yu2016} for efficient heating of hydroxyl groups within the nanofiber.
Additionally, both before and after fabrication and laser heating, the transmission spectrum of the ONF was measured by replacing the input to the ONF with light from a super luminescent diode~(SLD) (SLD1410, Thorlabs), and measuring the output directly with an optical spectrum analyzer.

\begin{figure}[hbt!]
\centering
\includegraphics[width=\columnwidth]{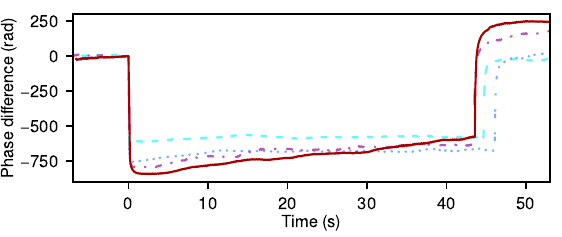}
\caption{Measured phase difference due to laser heating of the ONF at 5~mW (dashed cyan line), 10~mW (dotted blue line), 15~mW (purple dash-dotted line) and 20~mW (solid red line). In each case, the heating laser is turned on at 0~s, and turned off at $\sim45$~s. The corresponding phase shifts at the beginning of laser heating are $-650$, $-760$, $-800$, and $-850$~radians, respectively.}
\label{fig:phase}
\end{figure}
By guiding the 1390~nm heating laser into the ONF under vacuum, the temperature at the nanofiber section is increased, mainly due to absorption of the laser by the hydroxyl groups. The resulting increase in optical path length can then be inferred from the corresponding phase shift of the 852~nm probe light transmitted through the optical nanofiber.
Laser heating was performed at intensities of 5~mW, 10~mW, 15~mW, and 20~mW, with resulting phase shifts of approximately $-650$~rad, $-760$~rad, $-800$~rad, and~$-850$~rad, respectively, as shown in Fig.~\ref{fig:phase}. After heating the fiber at each power for approximately 45~s, the laser is turned off and the temperature of the nanofiber decreases, with corresponding phase shifts of 560~rad, 700~rad, 750~rad, and 830~rad, respectively. The decrease in magnitude of the phase shift compared to when the laser is turned on corresponds to an overall reduction in nanofiber temperature during laser heating, presumably due to desorption of hydroxyl groups. We also note that, while there appears to be a slow phase drift over longer time scales due to instability of the interferometer, the measured phase difference at short time scales is largely unaffected by this drift.

To estimate the temperature at the center of the ONF from the phase shift observed during laser heating, we employed a thermal model including heat conduction along the fiber axis, thermal radiation, and optical absorption caused by contaminants and OH groups attached to or incorporated in the optical nanofiber~\cite{Anderson2018, wuttke2013}. As laser heating of the ONF was performed under vacuum conditions, heat exchange from the nanofiber to any surrounding gas was neglected, as were temperature variations across the nanofiber cross-section.
The temperature distribution $T(z,t)$ along the optical nanofiber was calculated using the heat equation presented in previous works~\cite{Anderson2018, wuttke2013, Demtroder2003},
\begin{equation}
\label{eq:heat}
\begin{aligned}
c_\mathrm{p} \rho \partial_t T \pi a^2
&=
-\partial_z H_{\mathrm{rad}}(T)
+\partial_z H_{\mathrm{rad}}(T_0) \\
&\quad
+\lambda_\mathrm{c} (\partial_z^2 T)\pi a^2
+2\pi a k I(a).
\end{aligned}
\end{equation}
Here, $c_\mathrm{p}$ is the specific heat capacity of silica, $\rho$ is the density of silica, $a = a(z)$ is the position-dependent fiber radius, $\lambda_\mathrm{c}$ is the thermal conductivity of silica, $k$ is a proportionality constant determined by the surface absorber density and scattering cross section, and $I(a)$ is the optical intensity at the fiber surface.
The left-hand side of Eq.~(\ref{eq:heat}) represents the amount of heat absorbed per unit time by an infinitesimal volume $dV$ of the optical nanofiber, and by assuming a steady-state temperature distribution, this term can be set to zero.
The first and second terms on the right-hand side represent thermal radiation from the optical nanofiber and absorption of thermal radiation from the room-temperature environment at $T_0$, respectively. The third term represents heat conduction along the fiber axis, and the fourth term represents heat supplied by the heating laser.
The optical path length change $\Delta l$ caused by the temperature-dependent refractive index variation $n_\mathrm{eff}$ can be calculated as
\begin{equation}
\label{eq:pathlengthchange}
\Delta l
=
\int
\left[
n_{\mathrm{eff}}(T,z)
-
n_{\mathrm{eff}}(T_0,z)
\right]
dz,
\end{equation}
from which the relationship between the optical path length change and the phase shift $\Delta \phi$ is found to be
\begin{equation}
\label{sample:equation3}
\Delta \phi
=
\frac{2\pi}{\lambda}
\Delta l.
\end{equation}
The effective refractive index and thermal radiation model were solved following the method described in Ref.~\cite{Anderson2018}.
For the phase shifts due to heating shown in Fig.~\ref{fig:phase}, the corresponding center temperatures of the ONF were estimated to be 740~$^{\circ}$C (5~mW heating laser power), 840~$^{\circ}$C (10~mW), 870~$^{\circ}$C (15~mW), and 900~$^{\circ}$C (20~mW).

\begin{figure}[bt!]
\centering
\includegraphics[width=\columnwidth]{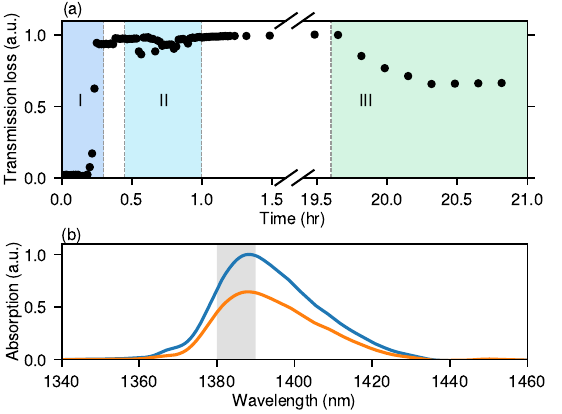}
\caption{(a)~Transmission loss in the wavelength range of 1380--1390~nm over the course of nanofiber fabrication and laser heating. Transmission loss has been scaled to the maximum loss that is observed due to water absorption after fabrication. The shaded regions correspond to (I) nanofiber fabrication, (II) installation into vacuum, and (III) laser heating. (b)~Example of the OH absorption spectra before and after laser heating (blue and orange lines, respectively). Gray shaded region indicates the integration area used to calculate transmission loss shown in (a).}
\label{fig:heating_laser}
\end{figure}
In order to quantify the reduction of OH groups due to heating, the transmission spectrum was measured during ONF fabrication and laser heating, as shown in Fig.~\ref{fig:heating_laser}(a). 
To ensure that we only consider transmission loss due to hydroxyl groups, we use the initial spectrum measured prior to nanofiber fabrication as a reference. We perform a least-squares baseline fit on subsequent spectra, excluding the region of OH-group absorption from 1360~nm to 1420~nm, and normalize these spectra to the initial reference data.
Following this correction, the intensity difference between the corrected spectrum and the reference spectrum was integrated over the wavelength range from 1380~nm to 1390~nm.
A strong increase in loss can be observed after the start of nanofiber fabrication, as highlighted by region~I in Fig.~\ref{fig:heating_laser}(a). The loss level remains constant during the installation into the vacuum chamber, indicated by region~II.
However, after the start of laser heating, indicated by region~III, the loss continuously decreased as the laser power was increased from 5~mW to 20~mW. An example of the OH absorption spectrum before and after laser heating is shown in Fig.~\ref{fig:heating_laser}(b). There is a clear decrease in absorption across the 1360--1420~nm region, showing that water is desorbed from the nanofiber during laser heating. We note that the peak of the absorption spectrum does not occur at exactly 1385~nm due to the spectral profile of the SLD, which has an intensity maximum at 1420~nm. The hydroxyl groups could not be fully desorbed due to the limits of our heating laser power, and we expect that further increasing the laser power may lead to more desorption. However, care must be taken not to exceed the glass transition temperature of silica. We also find that heating for extended periods of time at the same laser power does not noticeably desorb additional OH groups.

\begin{figure}[bt!]
\centering
\includegraphics[width=\columnwidth]{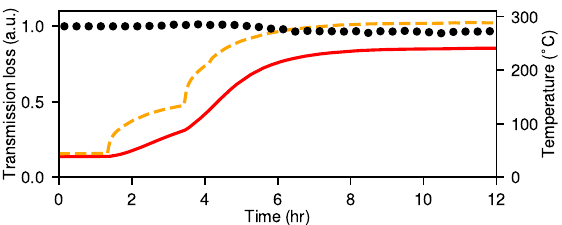}
\caption{Transmission loss in the wavelength range of 1380--1390~nm during baking of the nanofiber in vacuum (black circles, left axis). Orange dashed line and red line show the temperature at the chamber surface and nanofiber jig, respectively (right axis).
}
\label{fig:baking}
\end{figure}
For comparison with the laser heating method, we also perform experiments where the nanofiber is heated via thermal conduction by baking the vacuum chamber, using a ribbon heater wrapped around the outside of the chamber. During the baking experiment, the temperatures of the chamber surface and the jig holding the ONF inside the vacuum chamber were monitored using a pair of thermocouples. We note that it is not appropriate to use the interferometer to extract the nanofiber temperature here, as the phase shift solely due to nanofiber heating cannot be discerned from the overall change in optical path length of the entire heated fiber. We choose to use a jig made of Invar to hold the nanofiber, due to having a relatively low thermal expansion coefficient up to $\sim\!200~^\circ$C. As for the laser heating experiment, the spectrum measured at the beginning of the experiment was used as a reference, and the integrated intensity difference in the wavelength range from 1380~nm to 1390~nm was continuously monitored. The results are shown in Fig.~\ref{fig:baking}. After slowly increasing the temperature over six hours, the jig was maintained at approximately 240~$^{\circ}$C for another six hours, after which the nanofiber broke, most likely due to stresses caused by different rates of thermal expansion of the glue and the ONF jig compared to the fiber. Prior to the nanofiber breaking, the transmission loss had only improved by about 5\%, compared to around 40\% for the laser heating method. We have performed several further baking attempts with similar results. To prevent issues with breaking the nanofiber and to proceed further with the baking experiments, it may be possible to construct a jig from silica glass, as well as to fix the nanofiber to the jig using a glass epoxy. However, our initial results have highlighted the difficulty of significantly increasing the nanofiber temperature using this method, due to poor heat conduction through to the nanofiber in vacuum, and the impracticality of reaching temperatures as high as those achieved by guiding light within the fiber.

In this study, we investigated laser heating as a method for removing OH groups introduced during the fabrication process of optical nanofibers.
Changes in the loss in the wavelength range from 1380~nm to 1390~nm in the transmission spectrum of the optical nanofiber provided evidence for the desorption of OH groups induced by laser heating.
In addition, OH groups were also desorbed from the optical nanofiber by heating the vacuum chamber externally and transferring heat through thermal conduction. However, laser heating was found to be much more efficient.
These results demonstrate that the proposed method is a practical post-fabrication treatment for reducing absorption of light by OH groups in optical nanofibers without requiring modifications to the fabrication process. The method is expected to contribute to the realization of quantum networks operating at telecom wavelengths based on CQED systems using Yb atoms and optical nanofiber cavities.

{\bf Funding} This work was supported by JST ASPIRE Grant Number JPMJAP2511, JST CREST Grant Number JPMJCR25I1, and JST Moonshot R\&D Grant Number JPMJMS256K, Japan.

{\bf Competing interests} T.A. is a co-founder of, and an equity shareholder in, Nanofiber Quantum Technologies, Inc.

\end{document}